\documentclass[twocolumn,amsmath,amssymb,prl]{revtex4}
\usepackage{latexsym}
\usepackage{graphicx}
\usepackage{psfig}
\usepackage{color}
\usepackage[colorinlistoftodos]{todonotes}
\usepackage{verbatim}
\usepackage{parskip}

\usepackage[letterpaper, margin=1in]{geometry}
\begin{document}

\title{Voltage-tunable superconducting resonators: a platform for random access quantum memory}
\author{Kasra Sardashti$^{1}$}
\author{Matthieu C. Dartiailh$^{1}$}
\author{Joseph~Yuan$^{1}$}
\author{Sean Hart$^{2}$}
\author{Pat Gumann$^{2}$}
\author{Javad~Shabani$^{1}$}

\affiliation{$^{1}$Department of Physics, New York University, NY 10003 USA
\\
$^{2}$ IBM Quantum, IBM T.J. Watson Research Center, NY 10598 USA
}
\date{\today}


\begin{abstract}
In quantum computing architectures, one important factor is the trade-off between the need to couple qubits to each other and to an external drive and the need to isolate them well enough in order to protect the information for an extended period of time. In the case of superconducting circuits, one approach is to utilize fixed frequency qubits coupled to coplanar waveguide resonators such that the system can be kept in a configuration that is relatively insensitive to noise. Here, we propose a scalable voltage-tunable quantum memory (QuMem) design concept compatible with superconducting qubit platforms. Our design builds on the recent progress in fabrication of Josephson field effect transistors (JJ-FETs) which use InAs quantum wells. The JJ-FET is incorporated into a tunable coupler between a transmission line and a high-quality resonator in order to control the overall inductance of the coupler. A full isolation of the high-quality resonator can be achieved by turning off the JJ-FET. This could allow for long coherence times and protection of the quantum information inside the storage cavity. The proposed design would facilitate the implementation of random access memory for storage of quantum information in between computational gate operations.
\end{abstract}

\pacs{}
\hspace{1pt}
\maketitle

\section{Introduction}
\label{sec:introduction}

Noisy intermediate-scale quantum (NISQ) era's leading hardware is mostly built on scalable cavity quantum electrodynamics (cQED) architectures using superconducting qubits. Quantum hardware with 50+ qubits are available in superconducting platform where the system takes the form of a quantum network with superconducting coplanar waveguide resonators (bus couplers and readout resonators) providing paths for microwave photons to indirectly interact with processing qubit nodes. In this scheme a resonator is used to exchange a photon with its adjacent qubits for readout or entanglement operations. Superconducting resonators are typically easier to fabricate and optimize their characteristics compared to qubits where there are many more complex factors. Consequently, there is a rich literature on the optimization of materials growth and resonator fabrication processes that recently lead to resonator lifetimes of a few milliseconds \cite{place_new_2020, kjaergaard_superconducting_2020, richardson_fabrication_2016, kamal_improved_2016}. This progress has made superconducting resonators emerge as not only an integral part of the qubit system, but also as memory for quantum information. For memory applications where maintaining long coherence times are necessary, a weakly coupled resonator-qubit network with fixed frequency $f$ and high quality factor, $Q = \frac{f}{\Delta f}$ can limit the speed of quantum operations (increasing the length of quantum gates). This issue may be addressed by integrating tunable inductive elements into the resonators in order to drive them in and out of tune with various components within the circuit \cite{palacios-laloy_tunable_2008}. We note that there is always trade-off between tunability and fixed operation of quantum circuits which can be quantified by Quantum Volume \cite{cross_validating_2019}.\\

Achieving tunable resonators is expected to have major impacts in several fields of quantum information processing and quantum communication. In the former, large arrays of qubits and resonators as quantum random access memory will be needed for short-term storage of the processed information \cite{hann_hardware-efficient_2019}. In the latter, flying qubits are envisioned to be sent over long distances with resonator-based quantum memories and qubit processors acting as nodes to boost the transfer efficiency in the form of quantum repeaters \cite{namazi_ultralow-noise_2017}.\\

Tunable resonant cavities have been demonstrated in the past decade by several groups including \cite{palacios-laloy_tunable_2008, sandberg_exploring_2009, krantz_investigation_2013, pierre_resonant_2019}. Nearly all of these designs include a superconducting quantum interference device (SQUID) placed in the resonator voltage node. When a small magnetic flux is applied the SQUID’s inductance and consequently the resonator’s total inductance varies. Tuning the inductance is equivalent to tuning the resonant frequency for the resonator. The magnetic field can be applied off-chip by a coil in the vicinity of the sample and on-chip via the so-called current flux line. However, either of those approaches requires application of current directly to or near the chip causing excess energy dissipation and flux noise in the qubits.

\section{Design of the Quantum Memory}
\label{sec:QuMem Design}

\begin{figure}[htbp]
\centering
   \includegraphics[scale=0.23]{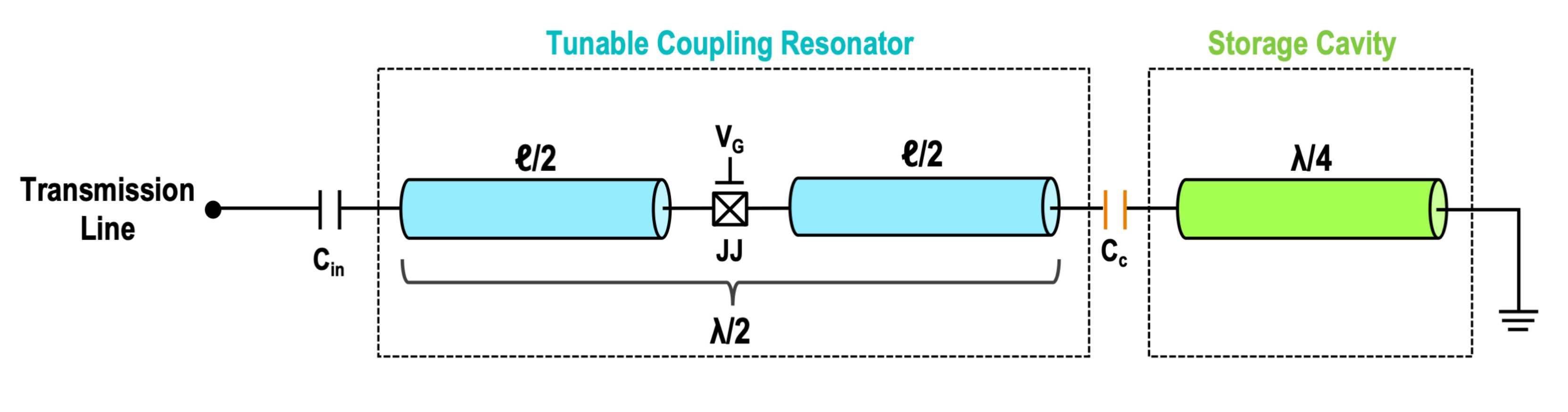}
    \caption{Schematic of the quantum memory device composed of a tunable coupling resonator and fixed-frequency storage resonator.}
    \label{fig.1}
\end{figure}

The recent development of epitaxial superconductor-semiconductor heterostructures \cite{shabani_two-dimensional_2016, krogstrup_epitaxy_2015} has allowed research into gate-tunable Josephson field effect transistors (JJ-FETs) on InAs near surface quantum wells \cite{mayer_superconducting_2019} and III-V nanowires \cite{casparis_superconducting_2018, luthi_evolution_2018}. Quantum devices fabricated on these materials exhibit record high junction transparencies and optimal inductance tunability via DC gate biasing \cite{shabani_two-dimensional_2016,kjaergaard_quantized_2016, kjaergaard_transparent_2017,mayer_superconducting_2019}. The JJ-FETs gate stacks are identical to room temperature III-V transistors where dielectrics can be deposited by atomic layer deposition of a thin layer of $Al_2O_3$ (or $HfO_2$) followed by evaporation of the metallic gate electrode (i.e. Ti/Au). Figure~\ref{fig.1} shows how the JJ-FET can be integrated into our resonator device for memory application. In this scheme, a $\lambda/2$ superconducting resonator is dissected into two sections of equal length by a JJ-FET. This resonant cavity labeled as “tunable coupling resonator (TCR)” performs the task of mediating the coupling between the transmission feed-line shown on the left and the $\lambda/4$ “storage cavity (SC)”, shown on the right. This way the SC can be isolated on demand. The SC is expected to have high internal quality factors while TCR can have high coupling to feed line for fast read/write. The TCR interfaces with the feed-line and storage cavity via two capacitors: input capacitor and coupling capacitor.\\

The envisioned operation of this device relies on DC voltage pulses applied to the gate electrode instead of current pulses applied to flux line when a SQUID is used as the tunable inductor \cite{pierre_resonant_2019}. Photons can be transferred from the TCR to the SC, and reciprocally, by tuning the JJ-FET critical current to achieve strong coupling between the two resonators for short periods of time, resulting in a photon swap. The coupling between the two resonators will determine the achievable read/write speed of the memory. The information stored in the SC can then be protected after each operation by driving the TCR out of tune.\\

\begin{figure}[htbp]
\centerline{\includegraphics[scale=0.6]{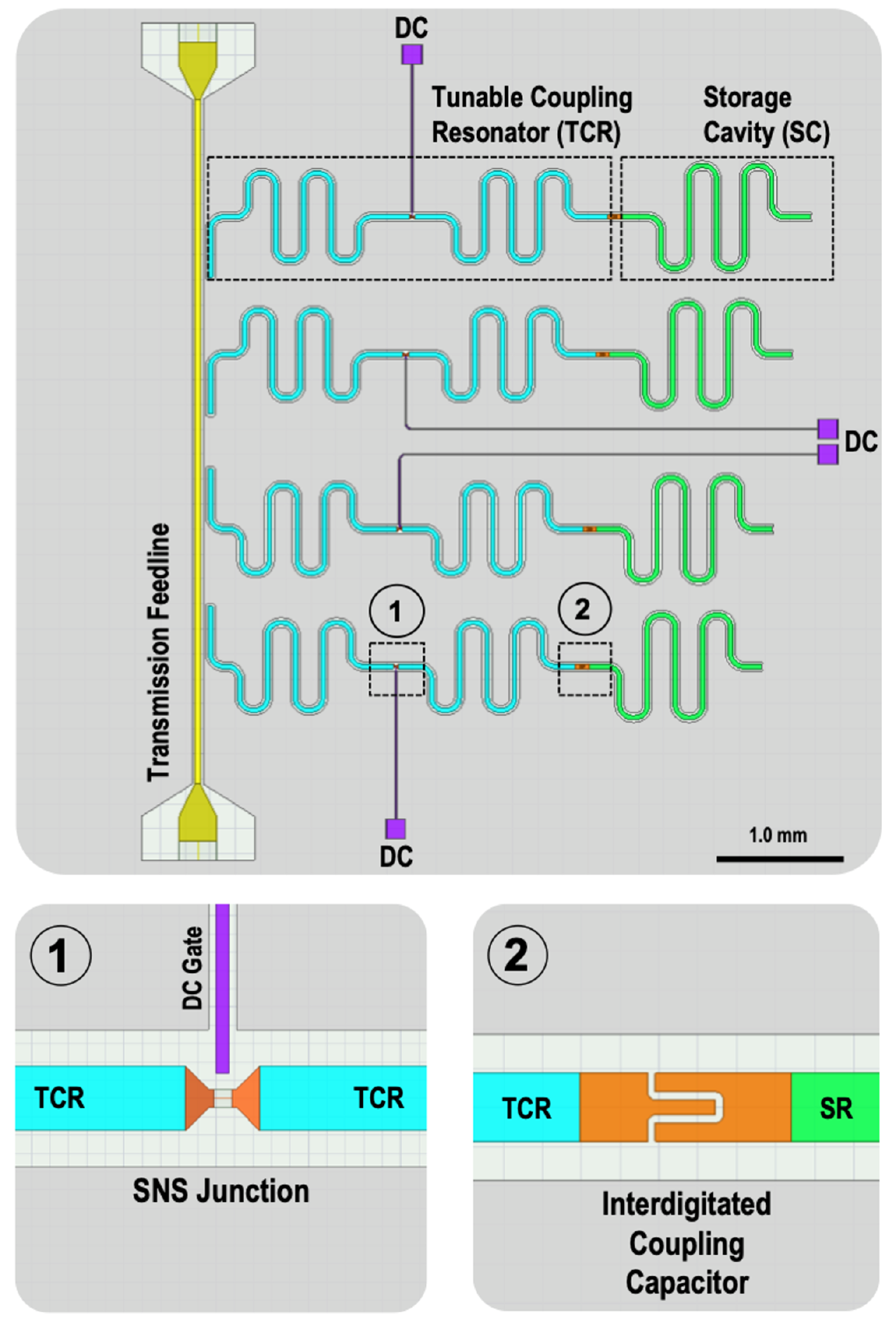}}
\caption{Top-down view of a model of the proposed QuMem design, used for HFSS simulations, with an array of four devices with characteristic resonance frequencies of 6.55 GHz, 6.65 GHz, 6.7 GHz and, 6.75 GHz for their storage cavities. The design includes a microwave transmission feedline (yellow), frequency-tunable coupling resonators (turquoise), DC bias lines for JJ-FET gates (purple) and fixed-frequency storage cavities (green).}
\label{fig.2}
\end{figure}

To optimize the planar microwave circuit for this application, electromagnetic modelling is carried out via a High Frequency Structure Simulator (HFSS) using Ansys 3D electromagnetic simulation software. Fig.~\ref{fig.2} displays a planar design for cQED quantum memories based on a hybrid superconductor-semiconductor platform. The schematic on the top is a top-down view of the design that includes four memory devices coupled to a transmission feed-line via tunable resonators. In this design, the coupling resonators are tuned by application of DC voltage pulses directly to the JJ-FET gate (section (1)). It should be noted that the schematics for the JJ-FETs are representative. The junction gap could also be covered with a high-k dielectric (i.e. $Al_2O_3$,  $Hf O_2$) and the gate electrode contacts (in purple). In our simulations the JJ-FETs are treated as lumped resistance-inductance-capacitance (RLC) elements. Section (2) of figure~\ref{fig.2} shows a magnified section of the design at the interface between the TCR and SC resonators. The coupling may be achieved via an interdigitated capacitor with only two teeth.\\

\begin{figure}[htbp]
    \centerline{\includegraphics[scale = 0.55]{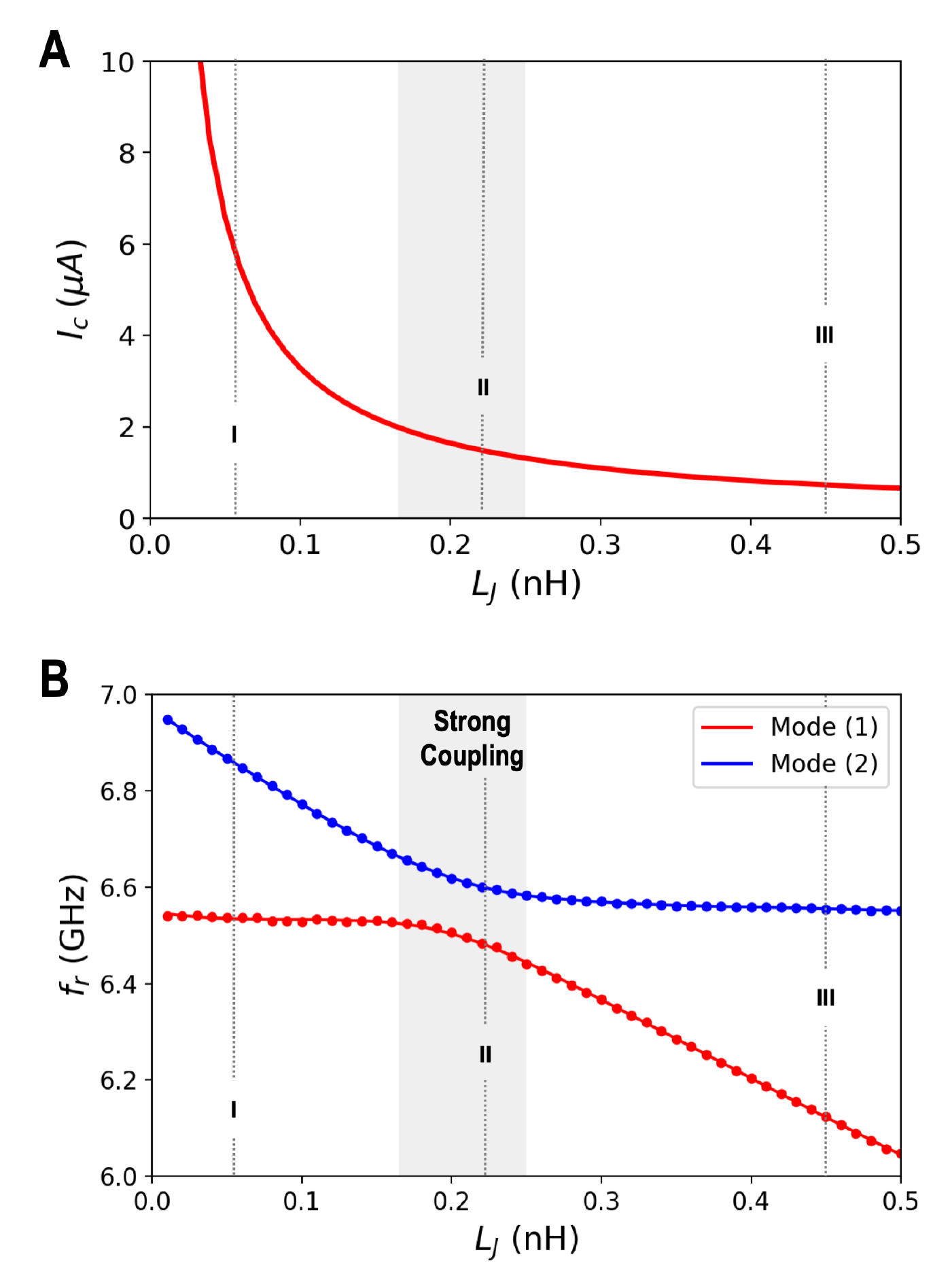}}
    \caption{(A) Critical current as a function of JJ-FET inductance following Eq.(1) (B) Resonant frequencies for mode 1 and mode 2 of the resonances, as functions of the JJ-FET inductance. Modes 1 and 2 represent the TCR and SC, respectively. As highlighted by the gray region, strong coupling between the two resonators occurs between 175 and 250 pH. The lines I, II and III denote the three inductance values at which the electrical field distribution maps in Fig.\ref{fig.4} (A) – (C) are calculated.}
    \label{fig.3}
\end{figure}

In order to simulate the TCR via HFSS, the overall inductance of the Josephson junction was varied over a wide range from 10 to 500 pH. This range is consistent with the critical current reported for InAs JJ-FETs (e.g. ref. \cite{mayer_superconducting_2019}) following the equation:

\begin{equation}
L_J=  \frac{\Phi_0}{2\pi I_c cos(\phi)}
\end{equation}

, where $\Phi_0$ is the flux quantum, $I_c$ is the critical current of the Josephson junction, and $\phi$ is the superconducting phase. This relationship at $\phi$ = 0 (a reasonable value for the TCR linear response regime) is plotted in Fig.~\ref{fig.3}(A) for typical values of $I_c$ in InAs JJ-FETs. The maximum achievable supercurrent can be tailored by the junction geometry as in epitaxial Al/InAs JJs the product of supercurrent and normal resistance reaches the Al superconducting gap ($I_{c}R_{n} \sim \Delta$). To target a particular maximum $I_{c}$, the $R_{n}$ can be adjusted by controlling the width of the junction.\\

Fig.~\ref{fig.3}B shows the resonant frequency calculated by HFSS for TCR (mode 1) and SC (mode 2) in the topmost resonator device in Fig.~\ref{fig.2} as a function of the JJ-FET inductance. This frequency range is particularly important as the frequency anti-crossing occurs between the two modes: a signature of strong coupling between the two superconducting resonant cavities which will be used to perform photon swap between the TCR and SC. The coupling strength can be estimated to be a few hundreds of MHz. This level of interaction between the TCR and SC offers a clear path to an efficient read/write process for the proposed memory device.\\

\begin{figure}[ht!]
    \centerline{\includegraphics[scale=0.6]{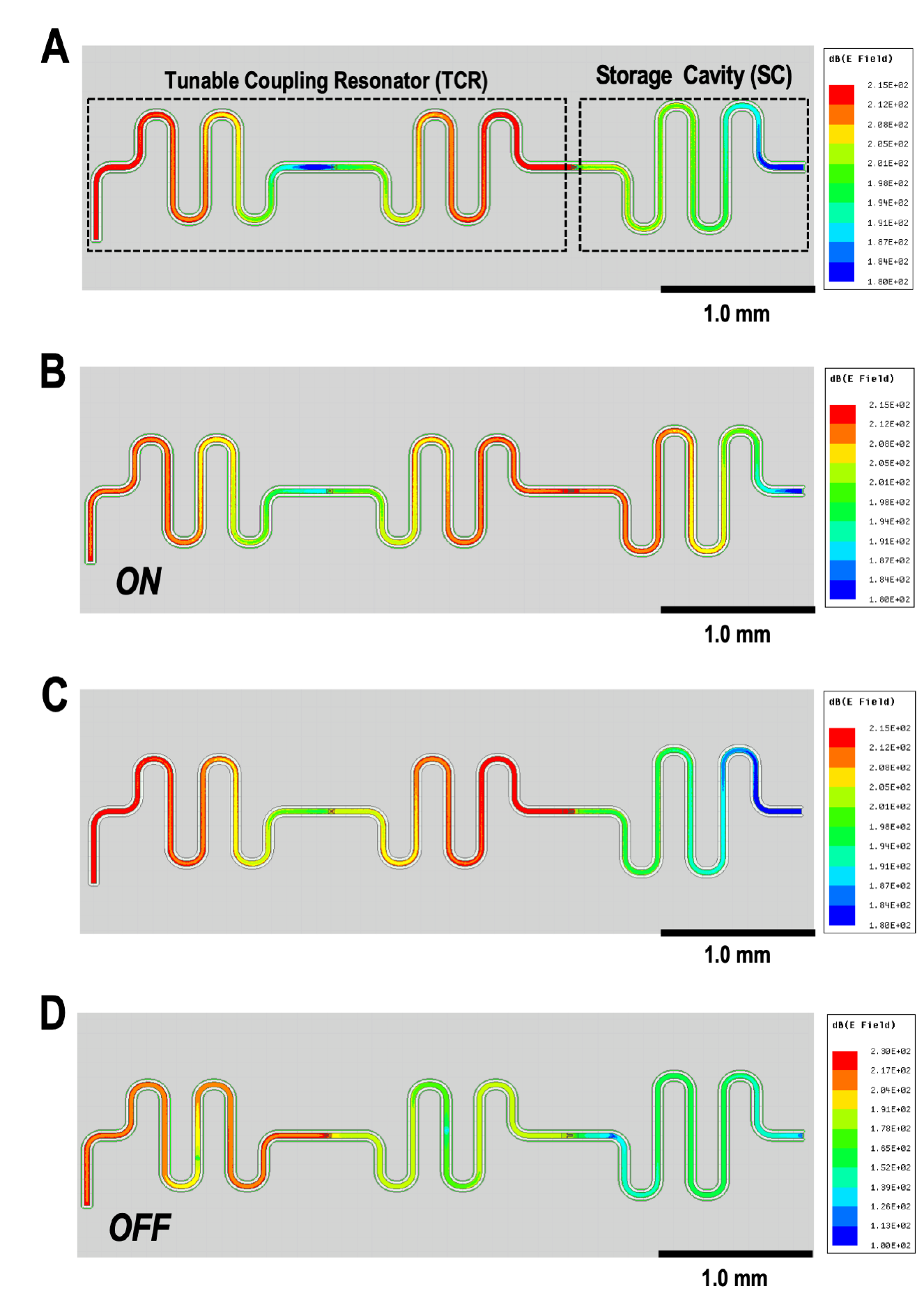}}
    \caption{Magnitude of electric field distributed between TCR and SC for four different conditions: (A) L = 50 pH, where TCR and SC are detuned; (B) L = 220 pH, where strong TCR-SC coupling occurs; (C) L = 450 pH, where TCR and SC are once again detuned; (D) Resistive mode for the tunable JJ-FET's fully depleted regimes in which TCR is split into two $\lambda/2$ resonators with frequencies of $\sim$ 13GHz.}
    \label{fig.4}
\end{figure}

The interaction between the resonant cavities could also be evaluated qualitatively using the electric field (E-field) mapping tool. The simulated maps for the magnitude of the electric field in dB are shown in Fig.~\ref{fig.4}, for four different conditions of the JJ-FET. The maps in Fig.~\ref{fig.4}A – C were calculated at inductance values denoted by vertical lines in Fig.~\ref{fig.3}B.  In Fig.~\ref{fig.4}A and 4C, JJ-FET inductance is at 50 pH and 450 pH, respectively, leading to weak interactions between the two cavities with the E-filed being concentrated mostly in the TCR with $\lambda/2$ mode. Once the inductance reaches the center of strong coupling region at about 220 pH (Fig.~\ref{fig.4}B), strong E-fields are observed across the interdigitated capacitor coupling the TCR to SC. This is consistent with a strong coupling of the electromagnetic field between the two cavities, which can be efficiently used for the read/write process.\\

A JJ-FET is only inductive up to the point where the critical current ($I_c$) is reached. By depleting the semiconducting weak-link of the junction, $I_c$ can be suppressed, therefore turning the junction into a resistive element with absolute resistance values in the order of few k$\Omega$s \cite{mayer_superconducting_2019}. To evaluate the behavior of our memory device in this resistive regime for JJs, we modelled our JJ-FET with a 1~k$\Omega$ resistor in between the two sections of the TCR. This is the lower limit for an Al/InAs JJ-FET, which could be achieved by tuning the design and materials properties of the junctions. Figure~\ref{fig.4}D displays the simulated E-field distribution between the TCR and SC for such configuration. Incorporation of 1~k$\Omega$  in the TCR leads to its splitting into two shorter $\lambda/2$ cavities with resonance frequencies of $\sim$ 13 GHz. Considering that our read/write frequencies are in the 6.5 GHz range, the resistive mode will be ideal for complete turn-off in the QuMem storage mechanism. Therefore, the JJ-FET resistive regime provides full protection of the storage cavity from decoherence in external coupling to the TCR.\\

\begin{figure}[ht!]
    \centerline{\includegraphics[scale=0.4]{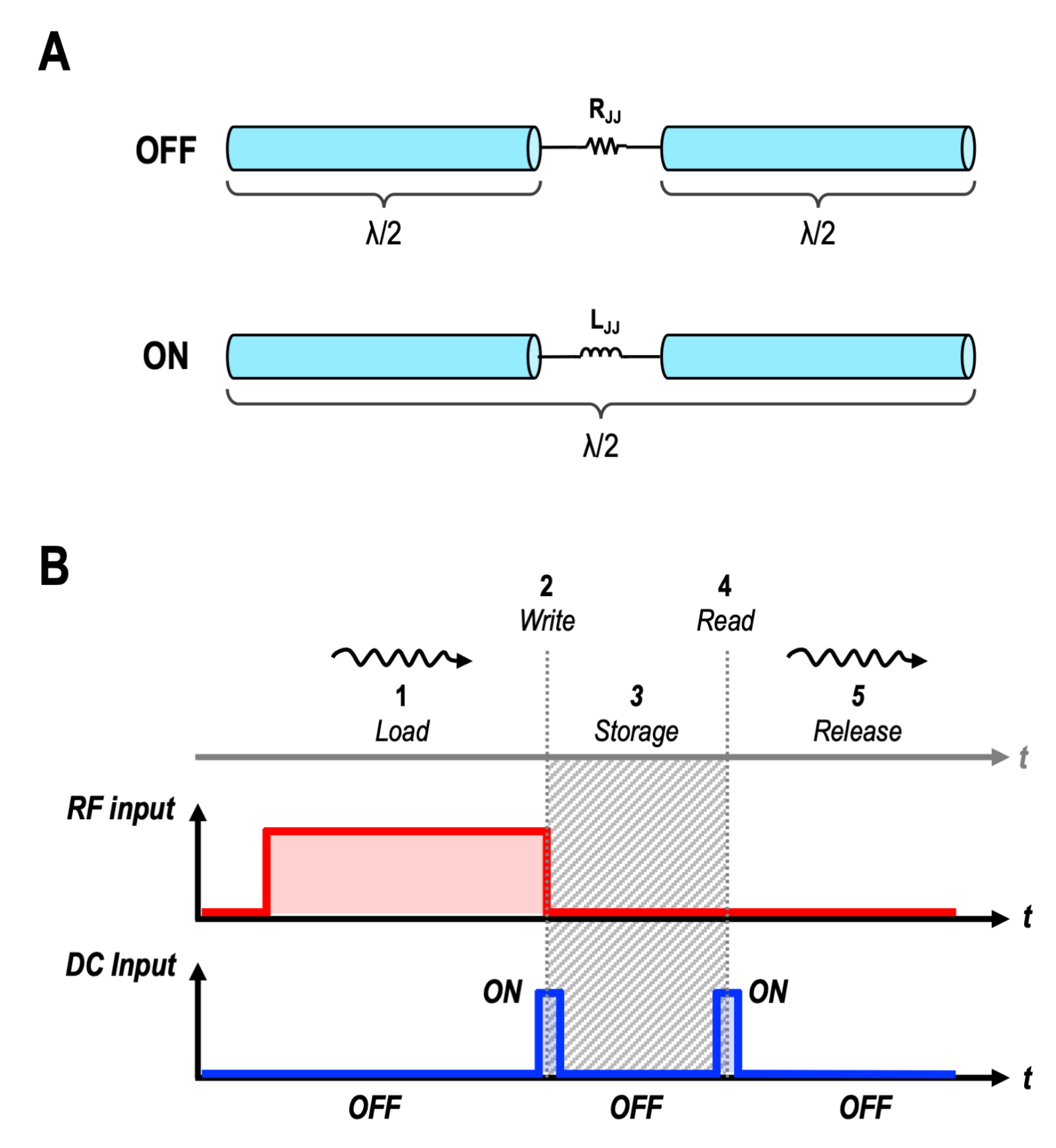}}
    \caption{The operation sequence for the proposed quantum memory device: (A) The state of the coupling resonator when the JJ is in complete depletion (OFF: equivalent to a resistor) or in accumulation (ON: equivalent to an inductor). The ON point corresponds to the voltage needed for the inductance to reach the level needed for strong coupling; (B) The pulse sequence for writing onto and reading from the quantum memory. The read and write processes occur as swap operation in the strong coupling regime. }
    \label{fig.5}
\end{figure}

 Based on the discussion above, there are two states of the JJ-FET employed in the memory operation process denoted by ``ON'' and ``OFF'' in Fig.~\ref{fig.5}A. In the ON state the JJ-FET acts as an inductive element with inductance values within the strong coupling regime (in this case $\sim$ 220 pH). In turn, in the OFF state the JJ-FET in full depletion acting as a resistive element with values $\geq$ 1 $k\Omega$.  Based on the proposed ON/OFF regimes, a pulse sequence for quantum information to be written onto and read from the QuMem is proposed in Fig.~\ref{fig.5}B. In this scheme the quantum information is carried by an radio frequency (RF) input pulse travelling within the transmission feedline. For most of the duration of this input RF pulse, the JJ-FET is in its resistive mode, therefore, the TCR and SC are in complete isolation. However, at the end of the RF pulse, a short DC pulse is applied to the JJ-FET gate in order to put TCR and SC in strong coupling with one another for exchanging information through SWAP operation. Afterwards, the exchanged information is securely stored in the SC in time scales defined by its quality factor, with TCR once again being fully decoupled in OFF state. This is equivalent to a write process onto a classical memory cell. For reading the quantum information from the QuMem, TCR and SC are once again placed in strong coupling regime with a short DC pulse.

\section{Scalability}

\begin{figure}[htbp]
    \centerline{\includegraphics[scale=0.2]{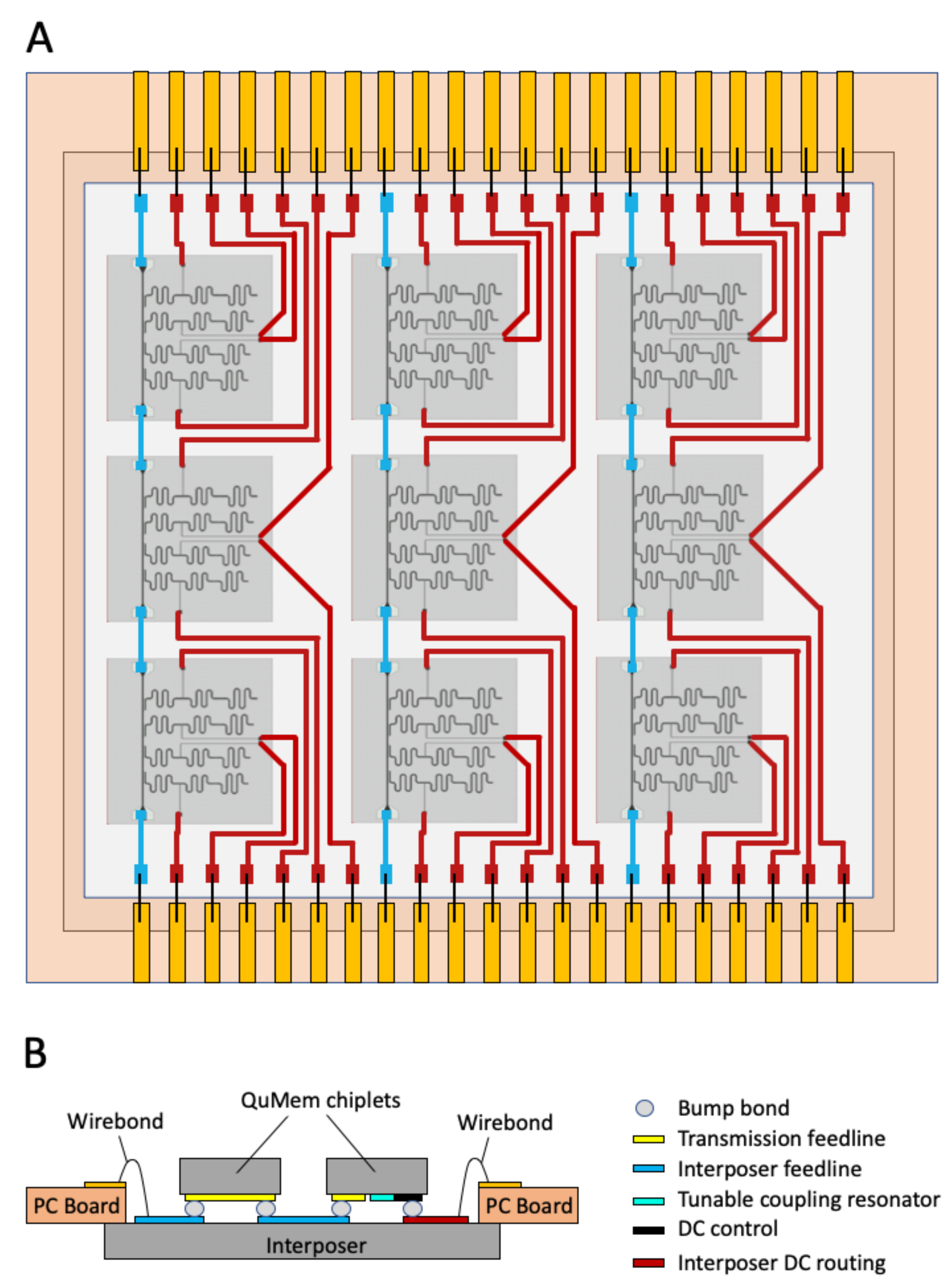}}
    \caption{Conceptual schematic of multi-chip quantum memory utilizing wire bonding or bump bonding technologies. Multiple chips can be added to an interposer composed of silicon or PCB/laminate, to increase the size of QuMem. Connectivity between chips and interposer can be accomplished by indium bump bonding methods.}
    \label{fig.6}
\end{figure}

The current quantum technology landscape is growing very rapidly. There are many research groups and industrial teams pursuing a wide variety of approaches, including superconducting qubits, solid state defects, entangled photons, and ion trap platforms \cite{gambetta_building_2017}. We believe that the next significant step in the development of quantum technologies is to demonstrate a versatile quantum memory, where solutions such as QuMem could be universally used across various platforms. As the number of qubits in quantum processors grows, so will the number of QuMem elements. Here, we propose a proven bump bonding approach for scaling up the quantum memory architecture, in the form of multiple chips bonded together by means of an interposer composed of silicon or PCB/laminate (see Fig.~\ref{fig.6}). State-of-the-art indium bump bonding techniques,  utilizing 200 mm silicon wafer technology, allow for connectivity of up to potentially 100 chips \cite{foxen_qubit_2017}. Alternative wire bonding methods can be applied as well, although those are typically less reliable and more time-consuming.
This framework allows for standard and established quantum characterization protocols to be applied \cite{abdelhafez_universal_2020, sheldon_characterization_2017}, enabling fast-turnaround characterization of individual QuMem chips in order to select the best-performing candidates. This modular solution enables faster screening of QuMems, maintenance/replacement of bad performers, cost reduction and cryogenic compatibility. This is particularly important as larger silicon devices have been demonstrated to have limited in their operations due to low thermal conductivity at milli-Kelvin temperatures \cite{white_thermal_1956}.

\section{Conclusion}
In this article, we propose a design concept for scalable voltage-tunable quantum memory. The resulting memory modules can be integrated with cQED superconducting processors or other qubit platforms, if efficient transduction is possible.  This design uses a gate-tunable hybrid JJ-FET as the main element to tune the resonance frequency of a superconducting resonator. This voltage-tunable resonator would control the coupling between a feedline and a superconducting storage cavity. Through HFSS simulations, we demonstrated that strong coupling can be achieved within a typical range of critical currents for hybrid gate-tunable Al-InAs Josephson junctions. When entering the JJ-FET's resistive regime, complete decoupling and isolation for the storage cavity may be achieved. This could lead to fully protected quantum information whose lifetime is defined by the quality factor of the storage cavity.\\

\section*{Acknowledgment}

NYU acknowledges partial support from Army Research Office grant number W911NF1810067 and ARO grant number W911NF1810115.   J. Y. acknowledges funding from the ARO/LPS QuaCGR fellowship reference W911NF1810067.  P.G. acknowledges this material is based upon work supported by the National Science Foundation under Grant No. 1839199.
This work is also supported in part by IBM Quantum, under Q Network for Academics program.



\section*{References}

\bibliography{IEEE_QuMem}

\end{document}